\documentclass[aps, prb, twocolumn]{revtex4-1}

\usepackage{enumerate}
\usepackage{comment}
\usepackage{amsmath}
\usepackage{amssymb}
\usepackage{color}
\usepackage[dvipsnames]{xcolor}
\usepackage{tikz}
\usepackage{amsthm}
\usepackage{graphicx}
\usepackage{pict2e}
\usepackage{bbold}
\usepackage{empheq}
\usepackage{subfigure}
\usepackage{xfrac}
\usepackage{float}

\usepackage{bm}
\usepackage{hyperref}
\hypersetup{
    colorlinks,
    linkcolor={red!50!black},
    citecolor={blue!50!black},
    urlcolor={blue!80!black}
}

\newcommand{\eqn}{Eq.~}

\newcommand{\fig}{Fig.~}
\newcommand{\figs}{Figs.~}

\newcommand{\editAK}[1]{{\color{black} #1}}
\newcommand{\editKZ}[1]{{\color{black} #1}}

\begin{document}

\title{Semimetallic and semiconducting graphene-hBN multilayers with parallel or reverse stacking.}
 
\author{Xi Chen$^{1,2}$}
\author{Klaus Zollner$^{3}$}
\author{Christian Moulsdale$^{1,2}$}
\author{Vladimir I.~Fal'ko$^{1,2,4}$}
\author{Angelika Knothe$^{1,3}$}

\affiliation{$^1$National Graphene Institute, University of Manchester, Manchester M13 9PL, United Kingdom}
\affiliation{$^2$Department of Physics and Astronomy, University of Manchester, Oxford Road, Manchester, M13 9PL, United Kingdom}
\affiliation{$^3$Institut f\"ur Theoretische Physik, Universit\"at Regensburg, D-93040 Regensburg, Germany}
\affiliation{$^4$Henry Royce Institute for Advanced Materials, University of Manchester, Manchester, M13 9PL, United Kingdom}
\date{\today}


\begin{abstract}
{We theoretically investigate  3D layered crystals of alternating graphene and hBN layers with different symmetries. Depending on the hopping parameters between the graphene layers, we find that these synthetic 3D materials can feature semimetallic, gapped, or Weyl semimetal phases. Using first-principles calculations to parameterize the low-energy Hamiltonians we establish the most likely electronic phases. Our results demonstrate that 3D crystals stacked from individual 2D materials represent a synthetic materials class with emergent properties different from their constituents. }
\end{abstract}
 
\maketitle
\section{Introduction}
Thanks to the recent progress in the layer-to-layer assembly of two-dimensional atomic lattices, it is now possible to combine individual atomic layers to create advanced synthetic crystals that would be difficult to achieve with any other bottom-up technique. Such layered three-dimensional (3D) materials with engineered stacking series can exhibit emergent characteristics different from the properties of their individual constituent layers. Moreover, such assembly of layers allows for multiple stacking orders of consecutive layers with different symmetries. Therefore, 3D  crystals obtained from stacking individual atomic layers one by one represent a synthetic materials class compared to the individual 2D sheets and their few-layer counterparts \cite{doi:10.1126/science.aba1416}.

One widespread choice is to combine graphene with hexagonal boron nitride (hBN). Heterostructures of various numbers and stacking arrangements of graphene and hBN layers feature, e.g., diverse super-lattice moir\'e effects \cite{wangCompositeSupermoireLattices2019,   moonElectronicPropertiesGraphene2014, sakaiElectronicStructureStability2011, yangSituManipulationVan, wangNewGenerationMoire2019, finneyTunableCrystalSymmetry2019, wallbankMoireSuperlatticeEffects2015,  zollnerHeterostructuresGrapheneHBN2019}, topological states \cite{huMoireValleytronicsRealizing2018, moulsdaleKagomNetworkChiral2022}, correlated states and superconductivity \cite{chenSignaturesTunableSuperconductivity2019,    sunCorrelatedStatesDoublyaligned2021}, dielectric and ferroelectric properties \cite{https://doi.org/10.1002/smll.201501766, https://doi.org/10.48550/arxiv.2102.12398, https://doi.org/10.48550/arxiv.2209.10636}, and exotic Hofstadter butterflies \cite{chenZeroenergyModesValley2016, fabianWannierDiagramBrownZak2021}.

Here, we provide both hybrid tight-binding-$k.p$-theory and density functional theory (DFT) calculations for the low-energy states of 3D synthetic crystals constructed from alternately stacked graphene and hBN monolayers. At a single interface between graphene and hBN monolayers, the two lattices have slightly different lattice constants, and straining one lattice to fit the lattice constant of the other is energetically very costly \cite{unpublished}. However, in a 3D bulk system with hBN layers alternating on either side of each monolayer graphene, the adhesion energy would promote the favourable atomic stacking of carbon and boron/nitrogen atoms. In compliance with recent first-principles  \cite{giovannettiSubstrateinducedBandGap2007} and diffusion Monte Carlo calculations \cite{unpublished} our DFT results confirm  the interplay of adhesion and strain to favour carbon atoms to  align with boron atoms to minimize the total energy \cite{lebedevaComparisonPerformanceVan2017, sakaiElectronicStructureStability2011}.

\begin{figure}[t]
    \centering
    \includegraphics[width=1\linewidth]{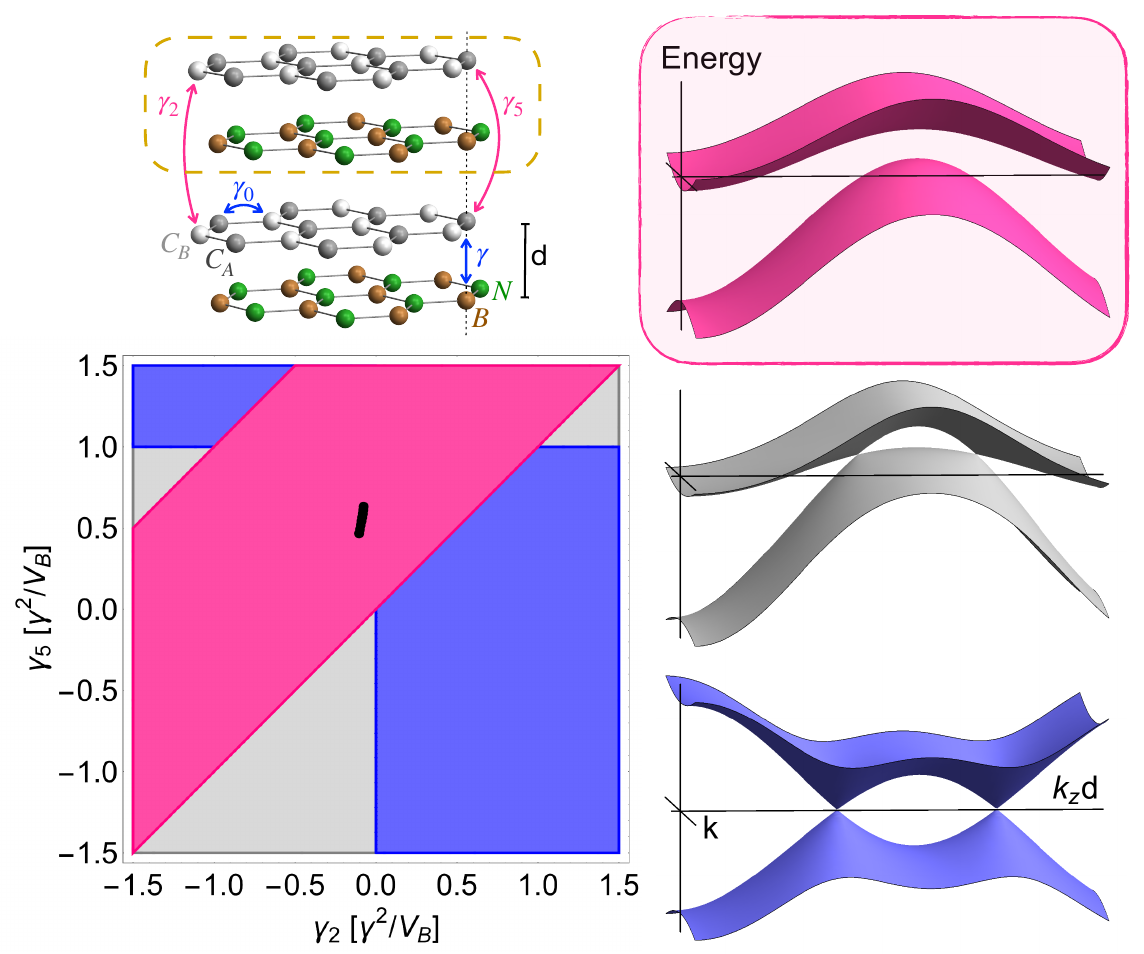}
    \caption{Possible electronic phases for 3D crystals of graphene and hBN monolayers periodically stacked in the z-direction (sketch on the top left, where the dashed orange line indicates the unit cell). Depending on the  hopping parameters $\gamma_5$ and  $\gamma_2$ between graphene sheets, this artificial material may exhibit different semimetal phases illustrated by the  dispersion sketches. The black mark in the parameter map indicates parameter values we obtain from DFT calculations yielding the range $\gamma_5 \in \{ 0.48 \,\gamma^2/V_B , 0.64  \,\gamma^2/V_B \}$ and  $\gamma_2 \in \{ -0.10 \,\gamma^2/V_B , -0.07  \,\gamma^2/V_B \}$ with increasing interlayer distance d, identifying overlapping bands as the most likely phase (pink regime).}
    \label{fig:1}
\end{figure}

We study   periodic 3D stacking obtained by  translating the hBN layers in the stacking process (hence all hBN layers are parallel to each other, c.f.~\fig{}\ref{fig:1}). 
For this perfectly z-periodic  Gr/hBN stack we study the resulting 3D band structures.

In a formal parametrization of the low-energy Hamiltonian, we find that, depending on the inter-layer graphene hopping parameters, such a 3D crystal can feature different types of semimetallic spectra, including overlapping electron and hole pockets, as well as type I and type II Weyl cones. Such Weyl semimetals are 3D phases of matter whose electronic properties and topology entail  protected surface states and anomalous responses to external electric and magnetic fields \cite{armitageWeylDiracSemimetals2018,chenTopologicalHybridSemimetal2021,burkovWeylMetals2018, PhysRevB.105.125131, PhysRevB.105.235408, https://doi.org/10.48550/arxiv.2209.14331, https://doi.org/10.48550/arxiv.2301.02965}.
Subsequently, we use DFT calculations to obtain estimates for the coupling parameters and establish the actual phase of the 3D Gr/hBN crystal to feature overlapping, non-topological bands, cf.~\fig{}\ref{fig:1}.

This manuscript is structured as follows: In sections \ref{sec:Heff} and \ref{sec:Bands}, we discuss the low-energy effective Hamiltonian for 3D Gr/hBN stacks with parallel hBN layers yielding the phase diagram of  \fig{}\ref{fig:1}. In section \ref{sec:DFT} we present our DFT calculations used to parameterize the hopping parameters and to determine the most likely phase. In section  \ref{sec:AP} we  consider an alternative stacking sequence,  where the hBNs are alternatingly rotated before placing them onto the graphene, resulting in adjacent hBN crystals in every second layer being antiparallel to each other (cf.~\fig{}\ref{fig:3}). For this case of antiparallel stacking of hBN layers, we again analyse the resulting 3D band structures in terms of the low-energy Hamiltonian and complimentary DFT calculations. 
We discuss our results in section \ref{sec:V} and give details of the derivations and DFT calculations in the appendices.

\section{Parallel stacking}

\subsection{Phenomenological low-energy effective Hamiltonian}
\label{sec:Heff}

Starting from a hybrid $k.p$ theory tight-binding approach\cite{garcia-ruizFullSlonczewskiWeissMcClureParametrization2021} for the  3D graphene/hBN crystal we derive the low-energy effective Hamiltonian for the electrons on the graphene layers subject to perturbations from the adjacent hBNs \cite{moonElectronicPropertiesGraphene2014, mccannLandauLevelDegeneracyQuantum2006,giraudpaulStudyElectronicStructure2012}. Hybridization between graphene and hBN orbitals has been studied previously and used in earlier studies of, e.g., moir\'e superlattices of single Gr/hBN interfaces \cite{wallbankMoireSuperlatticeEffects2015}. Here, we use second order perturbation theory in the interlayer hoppings to exclude the boron and nitrogen bands (see Appendix \ref{sec:AppA} for details of the calculation).  For the 3D Gr/hBN stacks with translated (parallel, p) hBN layers as depicted in  \fig{}\ref{fig:1}, the resulting low-energy Hamiltonian for the electrons in graphene reads,
\begin{widetext}
\begin{equation}
H_{\mathrm{p}}=\begin{pmatrix}
 -\frac{2\gamma^2}{V_B}(1+\cos (2 {k_z d}))+2\gamma_5\cos (2 {k_z d}) & v\pi^{\dagger}\\
 v\pi & 2\gamma_2\cos (2 {k_z d})
\end{pmatrix},
\label{eqn:Heffpar}
\end{equation}
\end{widetext}
operating in the space spanned by the two-component wave function $\Psi=(\psi_{C_A},\psi_{C_B})$ describing electronic amplitudes on the $C_A$ and $C_B$ sites of the graphene lattice. Further, $\pi=p_x+i p_y$ ($\bm{p}=-i\hbar\nabla$), $V_B$ 
is the onsite potential of boron 
measured with respect to the on-site potentials of carbon, $v=\frac{\sqrt{3}a}{2}\gamma_0$ is the Fermi velocity, and 
$d$ is the distance between graphene and hBN as indicated in \fig{}\ref{fig:1}.

For a faithful description of low-energy features in the electronic structure it is crucial to retain all the relevant couplings between different atomic sites \cite{mccannLandauLevelDegeneracyQuantum2006}. Here, we take into account $\gamma$ (between carbon and boron atoms), as well as the inter-layer coupling parameters between graphene layers, $\gamma_5$, and $\gamma_2$ between the in-equivalent carbon atoms $C_A$ (separated by a boron atom) and $C_B$ (separated by a hollow position). The precise values of these hoppings are a priori unknown. To explore the full parameter space, we first treat $\gamma_5 $, and $\gamma_2$  as free parameters in relation to $\gamma $ 
in the following discussion of possible phases. Then, in section \ref{sec:DFT}, we use DFT to estimate these parameters and establish the actual phase.

 \begin{figure*}[htbp]
    \centering
    \includegraphics[width=1\linewidth]{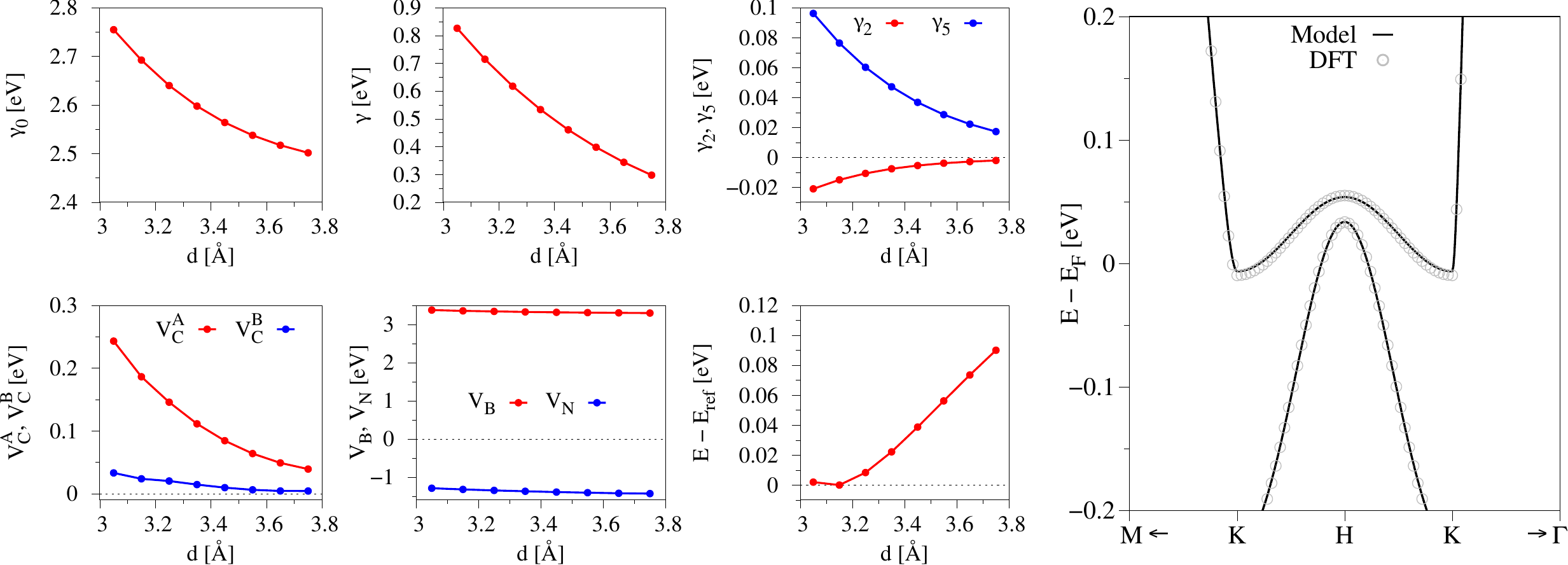}
    \caption{Parameters obtained by fitting the model Hamiltonian  (see Appendix \ref{sec:AppB}, \eqn{}\ref{DFT_Ham})  to the DFT-calculated low energy band structure, as function of the layer spacing $d$. In the rightmost figure, we explicitly compare the DFT and model low energy band structure for the lowest energy distance $d = 3.15$~\AA. } 
    \label{fig:DFT}
\end{figure*}

\subsection{Semimetal  band structures}
\label{sec:Bands}

The relative magnitude and sign of the hopping parameters between different atomic lattice sites determines the electronic properties of the 3D Gr/hBN crystals in \fig\ref{fig:1}.

We find that a 3D Gr/hBN crystal with parallely oriented hBN layers  either features overlapping electron and hole pockets, or type I, or  type II Weyl points depending on the hopping parameters $\gamma_2$ and $\gamma_5$. Figure~\ref{fig:1} demonstrates this parametric dependence of the electronic properties, showing the phase diagram in the plane spanned by the inter-layer hoppings and examples for the distinct possible 3D band structure types that we obtain from diagonalising $H_{\mathrm{p}}$ in \eqn\ref{eqn:Heffpar}. In the gapless phases, linear Weyl nodes\cite{cayssolIntroductionDiracMaterials2013, burkovWeylMetals2018, chiuClassificationTopologicalQuantum2016, burkovWeylSemimetalTopological2011, armitageWeylDiracSemimetals2018, goerbigTiltedAnisotropicDirac2008} form at momentum points $\bm{k}_0=(0,0,k_{z0})$ with 
\begin{equation}
    k_{z0}=\pm\frac{1}{2 d}\arccos{\big[-\frac{\gamma^2}{\gamma^2+V_B(\gamma_2-\gamma_5)}\big]}.
\end{equation}
These touchings can be type I Weyl nodes (closed or point-like Fermi surfaces, blue phase in \fig  \ref{fig:1}) or type II Weyl nodes  (overlap between electron and hole bands leading to open Fermi surfaces, gray phase in \fig  \ref{fig:1}) \cite{armitageWeylDiracSemimetals2018, https://doi.org/10.48550/arxiv.2209.14331, goerbigTiltedAnisotropicDirac2008}, and we find them to be Chern-nontrivial with Chern numbers $C=\pm1$.

\subsection{Modelling of graphene-hBN multilayers with density functional theory}
\label{sec:DFT}
 \begin{figure}[b!]
    \centering
    \includegraphics[width=0.8\linewidth]{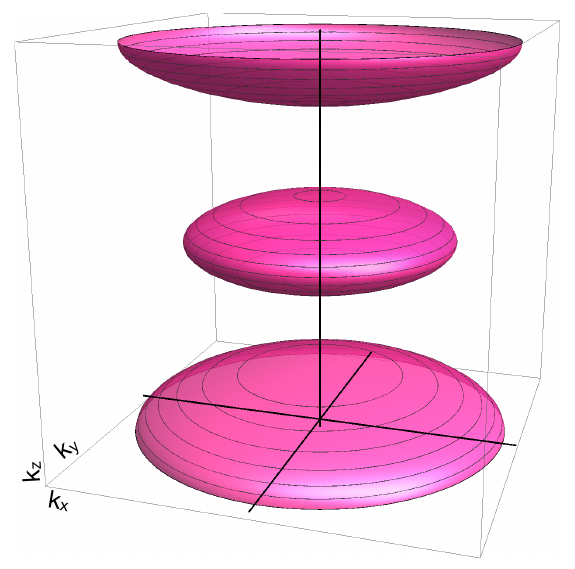}
    \caption{Fermi surface for $E_F=0$ of the overlapping bands of the parallely stacked Gr/hBN 3D crystals that we identify using the parameters obtained from DFT.} 
    \label{fig:PFermi}
\end{figure}
To estimate values for the parameters in \eqn \ref{eqn:Heffpar} we perform electronic structure calculations of the 3D Gr/hBN stack in \fig \ref{fig:1} by DFT~\cite{Hohenberg1964:PRB} with {\tt Quantum ESPRESSO}~\cite{Giannozzi2009:JPCM}. Self-consistent calculations are carried out with a k-point sampling of 36 × 36 × 18. We use an energy cutoff for charge density of 480 Ry and the kinetic energy cutoff for wavefunctions is 60 Ry for the scalar-relativistic pseudopotentials with the projector augmented wave method~\cite{Kresse1999:PRB} with the Perdew-Burke-Ernzerhof exchange correlation functional~\cite{Perdew1996:PRL}. Spin-orbit coupling is not included in the calculations. DFT-D2 van der Waals corrections are also included \cite{Grimme2006:JCC,Grimme2010:JCP,Barone2009:JCC}.

Our DFT calculations confirm that the atomic arrangement in \fig{}\ref{fig:1} is indeed the lowest energy configuration, where B atoms serve as dimer atoms for C atoms in graphene.
For this lowest-energy stacking, we fit the parameters of the model Hamiltonian.
We present the extracted values of the parameters as a function of the interlayer distance in \fig{}\ref{fig:DFT}. The black mark in the parameter plot of \fig{}\ref{fig:1} corresponds to these DFT parameter values in \fig \ref{fig:DFT}. The resulting DFT band structure for the lowest energy configuration yields an interlayer distance of $d = 3.15$~\AA{} and confirms the top right continuum model bands in \fig \ref{fig:1} (pink phase).  Based on these results we claim the most likely phase of the parallely stacked 3D Gr/hBN crystal to feature separated, overlapping bands. We show the corresponding shape of the $E_F=0$ Fermi surface in \fig{}\ref{fig:PFermi}. 

\editKZ{
Note that a small lattice mismatch of graphene and hBN does lead to moir\'{e} patterns
\cite{Jung2014:PRB,Moon2014:PRB,Argentero2017:NL}. Therefore, one cannot a priori exclude other stacking configurations within experimental setups. 
In Appendix \ref{sec:AppB}, we compare to the other commensurate high-symmetry stackings, where either N or both, N and B atoms, serve as dimer atoms. We also introduce model Hamiltonians, similar as above,  and parameterize them according to the DFT results.
We observe band touchings and conical band intersections for the parallel stacking configurations where graphene and nitrogen atoms dimerize, cf.~\figs{}\ref{fit_C_over_N} and \ref{fit_CC_over_BN} in Appendix~\ref{sec:AppB}.
Hence, the electronic phase is determined by the stacking configuration. Reducing the interlayer distance, achievable with external pressure, does not lead to a phase transition. We explicitly demonstrate this in the Supplemental Material, where we show movies of the evolution of the band structures as function of the interlayer distance for the different stackings. 
However, since all commensurate high-symmetry stackings are potentially present in realistic experimental structures, the full phase diagram in  \fig{}\ref{fig:1} is physically relevant, and one can expect local variations of the electronic phase in large scale samples.}

 \begin{figure*}[hbtp]
    \centering
    \includegraphics[width=1\linewidth]{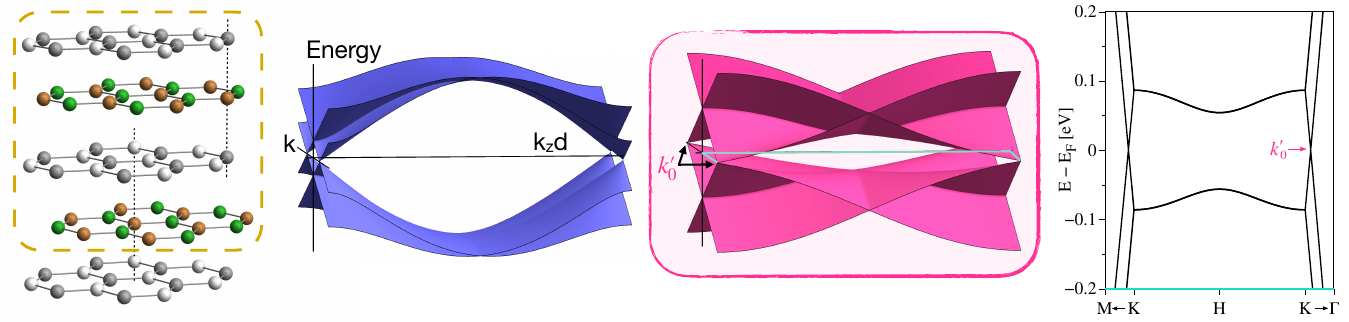}
    \caption{Possible dispersions for 3D Gr/hBN crystals with antiparallel stacking, where   adjacent hBNs in every second layer are rotated with respect to each other (sketch on the left). The  dashed orange box illustrates that the unit cell in the z-direction is twice as large for antiparallel stacking compared to parallel arrangements.  We chose similar parameters as for the phases (blue, magenta) in  \fig\ref{fig:1}. The phenomenological low-energy description in the magenta phase reproduces our  DFT results (rightmost dispersion for a cut along the turquoise path in momentum space where the two electron bands and the two hole bands are degenerate, respectively).
    }
    \label{fig:3}
\end{figure*}

\section{Antiparallel stacking}
\label{sec:AP}

We turn to 3D Gr/hBN crystals where the   adjacent hBN layers are rotated by 180$^{\circ}$ with respect to each other (antiparallel, ap, see the sketch in \fig{}\ref{fig:3}). In a similar fashion as for the parallel case we compute  the  low-energy Hamiltonian for  antiparallel stacking,
\begin{widetext}
\begin{align}
 H_{\mathrm{ap}}=\setlength\arraycolsep{-0pt}\begin{pmatrix}
 -\frac{2\gamma^2}{V_B}  & v\pi^{\dagger} &   e^{i 4 dk_z}\gamma_2 +\gamma_5 -\frac{\gamma^2}{V_B}  &0\\
v\pi &   -\frac{2\gamma^2}{V_B} &0 & \gamma_2 +  e^{i 4 dk_z} (\gamma_5-\frac{\gamma^2}{V_B})\\
  e^{-i 4 dk_z}\gamma_2 +\gamma_5 -\frac{\gamma^2}{V_B}  & 0 &-\frac{2\gamma^2}{V_B}  &v\pi^{\dagger}\\
0 & \gamma_2+e^{-i 4 dk_z} (\gamma_5-\frac{\gamma^2}{V_B}) &v\pi &-\frac{2\gamma^2}{V_B}\\
\end{pmatrix}.
\label{eqn:Heffalt}
\end{align}
\end{widetext}
 Note that the periodicity of the unit cell in the case of antiparallely stacked hBN layers, \fig{}\ref{fig:3}, is twice as large compared to the case of parallel stacking, cf.~\fig \ref{fig:1}. See Appendix \ref{sec:AppA} for details of the calculation.

We diagonalize $H_{\mathrm{ap}}$ in \eqn\ref{eqn:Heffalt} to obtain the 3D band structures of Gr/hBN stacks with antiparallel arrangement of adjacent hBN layers.  We compare different types of possible dispersions in \fig{}\ref{fig:3} for different values of the couplings $\gamma_5$ and $\gamma_2$ (similar values as in the parallel case, \fig{}\ref{fig:1}, for the blue and magenta phase). We find gapless band structures for all coupling parameters as a consequence of the restored inversion symmetry of the antiparallel stacking configuration compared to the parallel one (the gray and the pink phase in \fig{}\ref{fig:1} merge). The two electron bands and the two hole bands are degenerate, respectively, along $k_z$ with $k_x=k_y=0$. All bands show conical band touchings at zero energy along a ring of radius 
\begin{equation}
    k_{0}^{\prime}= \frac{1}{v}\big( \frac{\gamma^2}{V_B} -\gamma_2 -\gamma_5 \big),
\end{equation}
as well as for zero momentum at energies 
\begin{equation}\epsilon=\pm \frac{1}{V_B}\Big(\sqrt{(\gamma^2-(\gamma_2+\gamma_5)V_B)^2} -\gamma^2 \Big),
\end{equation}
in the conduction/valence band.

\begin{figure}[hbtp]
    \centering
    \includegraphics[width=0.8\linewidth]{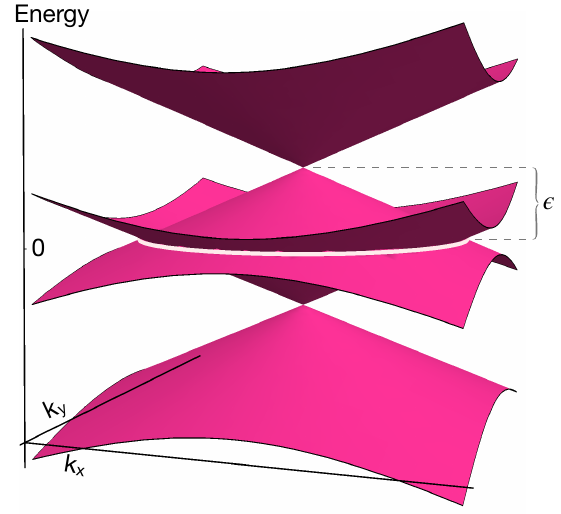}
    \caption{Conical crossings between the 3D bands of antiparallely stacked Gr/hBN stacks at $k_z=0$ that we determine for the parameters obtained from DFT. The white ring of radius $  k_{0}^{\prime}$ indicates the $E_F=0$ Fermi line. } 
    \label{fig:APFermi}
\end{figure}

 As in the case of parallely stacked crystals, our DFT calculations confirm the lowest-energy stacking configuration and the results of the continuum model with the coupling parameters of the magenta phase (rightmost panel in \fig{}\ref{fig:3}). We show the conical band touchings and the resulting $E_F=0$ Fermi line for the set of parameters from DFT in \fig{}\ref{fig:APFermi}.  

\section{Conclusion}
\label{sec:V}
 
We presented the possible electronic structure of 3D stacks of alternating graphene and hBN layers with different symmetries. The atomic arrangements we consider represent the most stable configurations for carbon and boron/nitrogen atoms in single adjacent layers\cite{giovannettiSubstrateinducedBandGap2007, unpublished} as we confirm using DFT calculations. Since the hopping parameters between graphene atoms in different layers are a priori unknown, we identify regimes with different electronic properties (semimetallic, overlapping bands, Weyl semimetals) upon varying these hoppings. We then extract estimates for the hopping parameters from DFT electronic structure calculations to determine the most likely electronic phases. \editAK{Regimes with different electronic band structures would make for distinctively different experimental signatures. Therefore, knowledge of the different possible types of dispersions presented here will help identifying signatures of the band structure in both transport and spectroscopy experiments to confirm the relative sign and magnitude of these  material parameters and set boundaries for their values.} Moreover, we anticipate that these out-of-plane hopping parameters could be manipulated, e.g., by applying perpendicular pressure to the 3D stacks \cite{szentpeteriTailoringBandStructure2021, pressure1, pressure2}. We hence confirm the stability of the electronic phase from DFT for different values  of the interlayer distance. 

Individual layers of graphene and hBN are very commonly combined in heterostructures with increasing precision and control, making the proposed crystals of alternating monolayers achievable in experiment. Using these 3D stacks of graphene and hBN as examples, we have demonstrated that artificial 3D crystals of individual atomic layers represent an advances synthetic 3D materials class with intriguing, potentially topologically non-trivial electronic properties only now achievable in experiments \cite{PhysRevResearch.2.022010, doi:10.1126/science.aba1416, doi:10.1021/acsnano.1c03698}. Besides the cases of alternating sequencing studied in this work, one may consider other stacking sequences with longer periods\cite{garcia-ruizSpectroscopicSignaturesElectronic2019, garcia-ruizElectronicRamanScattering2020}, stacking faults, interlayer twisting,  and the combination of multiple different 2D materials. Such considerations are left for further studies.

\section{Acknowledgements}
We thank Neil Drummond, Elaheh Mostaani, and Kostya Novoselov for insightful discussions. This study has been supported by the European Graphene Flagship Core3 Project,
 EPSRC Grants EP/W006502/1, EP/V007033/1, EP/S030719/1, EPSRC CDT Graphene-NOWNANO EP/L01548X/1, and by the Deutsche Forschungsgemeinschaft (DFG, German Research Foundation) SFB 1277 (Project No. 314695032), SPP 2244 (Project No. 443416183).
\clearpage

\appendix

\section{Derivation of the low-energy Hamiltonians}
\label{sec:AppA}

For the parallel stacking we start from the $4\times4$ Hamiltonian,
\begin{equation}
    H=\begin{pmatrix}
 H_{\mathrm{G}} &T^{\dagger}\\
T & H_{\mathrm{hBN}}
\end{pmatrix},
\label{eqn:HParDer}
\end{equation}
in the basis of the graphene and hBN  atomic sites $(C_A, C_B, N, B)$, where,
\begin {equation}
\begin {gathered}
H_{\mathrm{G}}=\begin{pmatrix}
V_{C}^A+2\gamma_5\cos ({2 k_z d})&v\pi^{\dagger}\\
v\pi &V_{C}^B+ 2\gamma_2\cos ({2 k_z d})
\end{pmatrix}, \\
H_{\mathrm{hBN}}=\begin{pmatrix}
V_N&0\\
0 & V_B
\end{pmatrix},\\
T = (1+e^{-i 2 k_z d}) T_0,\\
T_0 = \frac 13 \sum_{j = 0} ^ 2 e^{i (\bm{K}_j - \bm{K}_0) \cdot \bm{r}_0}
\begin{pmatrix}
    \gamma_N & \gamma_N e^{-i 2 \pi j / 3} \\
    \gamma_B e^{i 2 \pi j / 3} & \gamma_B
\end{pmatrix},\\
\bm {K}_j = \frac{4\pi}{3a} \bigg (\cos \frac {j 2 \pi} 3,\sin \frac {j 2 \pi} 3\bigg), \quad (j = 0, 1, 2)
\end {gathered}
\end {equation}
and for the relaxed equilibrium stacking considered in the main text the interlayer offset is $\bm{r}_0=(0,\frac{a}{\sqrt{3}})$. Eliminating the hBN sites,
\begin{equation}
    H_{\mathrm{p}}=H_{\mathrm{G}}+T^{\dagger}(-H_{\mathrm{hBN}})^{-1}T,
\end{equation}
 assuming $V_{C}^A, V_{C}^B \ll V_B$, and setting $\gamma\equiv\gamma_B$, we arrive at the expression in \eqn\ref{eqn:Heffpar} of the main text.

Similarly, for the alternative antiparallel stacking, we start from the Hamiltonian,
\begin{equation}
    \tilde{H}=\begin{pmatrix}
 H_{\mathrm{GG}} &\tilde{T}^{\dagger}\\
\tilde{T} & H_{\mathrm{hBNhBN}}
\end{pmatrix},
\end{equation}
in the basis of atoms on the upper and lower layers, $(C_A, C_B, \tilde{C}_A, \tilde{C}_B, N, B, \tilde{B}, \tilde{N})$, where
\begin{widetext}
\begin {equation}
\begin {gathered}
H_{\mathrm{GG}}=
\begin{pmatrix}
V_{C}^A&v\pi^{\dagger}&\gamma_5+e^{i 4 k_z d}\gamma_2&0\\
v\pi &V_{C}^B &0& \gamma_2+e^{i 4 k_z d} \gamma_5 \\
\gamma_5+e^{-i 4 k_z d}\gamma_2&0&V_{C}^A& v\pi^{\dagger}\\
0&\gamma_2+ e^{-i 4 k_z d}\gamma_5&v\pi&V_{C}^B
\end{pmatrix},\\
H_{\mathrm{hBNhBN}}=
\begin{pmatrix}
V_{N}&0&0&0\\
0&V_{B} &0& 0\\
0&0&V_{B}& 0\\
0&0&0&V_{N}
\end{pmatrix},\quad
\tilde{T}= \gamma
\begin{pmatrix}
0 & 0 & 0 & 0\\
1 & 0 & 1 & 0\\
0 & e ^ {- i 4 k_z d} & 0 & 1\\
0 & 0 & 0 & 0
\end{pmatrix},
\end {gathered}
\end {equation}
\end{widetext}
 Then, we obtain $H_{ap}$ in \eqn\ref{eqn:Heffalt} via the usual transformation,
\begin{equation}
    H_{\mathrm{ap}}=H_{\mathrm{GG}}+\tilde{T}^{\dagger}(-H_{\mathrm{hBNhBN}})^{-1}\tilde{T},
\end{equation}
for $V_{C}^A, V_{C}^B \ll V_B $.

\section{Details of the DFT calculations}
\label{sec:AppB}

\subsection{Structural Information}

\begin{figure*}[htb]
     \includegraphics[width=0.7\linewidth]{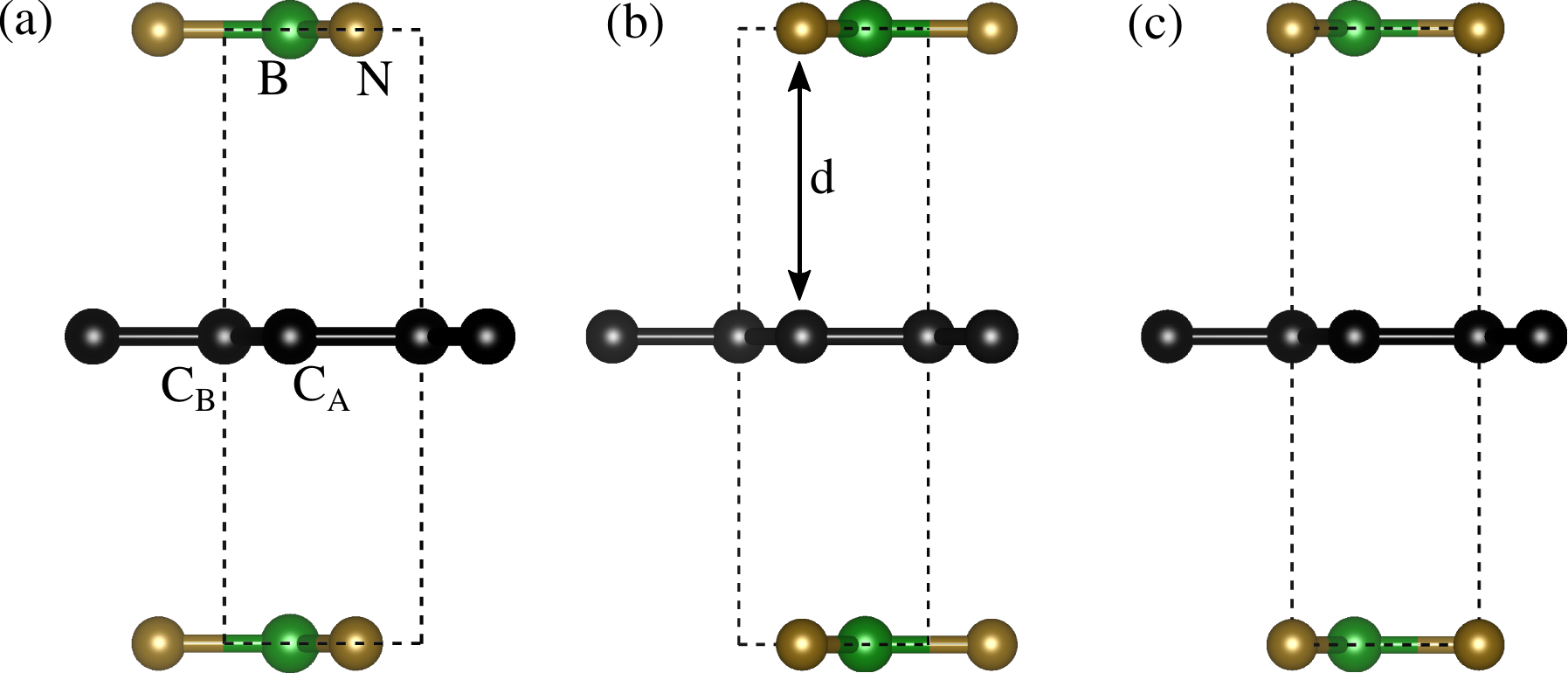}
     \caption{Geometries of the parallel stacking configurations. the dashed lines represent the unit cell. (a) C$_{\textrm{A}}$ and B are dimer atoms, while C$_{\textrm{B}}$ and N are non-dimer atoms. The interlayer distance $d = 3.35$~\AA. (b) C$_{\textrm{A}}$ and N are dimer atoms, while C$_{\textrm{B}}$ and B are non-dimer atoms. The interlayer distance $d = 3.50$~\AA. (c) C$_{\textrm{A}}$ and B, as well as  C$_{\textrm{B}}$ and N are dimer atoms. The interlayer distance $d = 3.55$~\AA.
     Initial interlayer distances from Ref.~~\cite{zollnerHeterostructuresGrapheneHBN2019}.
     }\label{Structure_parallel}
    \end{figure*}

Among the parallel configurations, we have three different choices to setup the unit cell~\cite{zollnerHeterostructuresGrapheneHBN2019}. 
The energetically most favorable is where C$_{\textrm{A}}$ and B are dimer atoms, while C$_{\textrm{B}}$ and N are non-dimer atoms, see Fig.~\ref{Structure_parallel}(a).
The in-plane lattice constant is chosen to be $a = 2.4846$~\AA, to compensate for the lattice mismatch between graphene and hBN~\cite{zollnerHeterostructuresGrapheneHBN2019}. Initial interlayer distances are summarized in the caption of Fig.~\ref{Structure_parallel}. We also perform an interlayer distance study to find the energetically most favorable situation.

\subsection{Model Hamiltonians to fit the DFT data}

The model Hamiltonian for the geometry in Fig.~\ref{Structure_parallel}(a) in the ordered basis (C$_{\textrm{A}}$, C$_{\textrm{B}}$, N, B) is

\begin{widetext}
\begin{flalign}
\label{DFT_Ham}
\mathcal{H}_{\textrm{p}}^{\textrm{(a)}} = & \begin{pmatrix}
V_C^A + 2\gamma_5\cos(k_z c) & \gamma_0 f(\bm{k}) & 0 & \gamma_1^B \left[1+\textrm{exp}(\textrm{i}k_z c)\right]  \\
\gamma_0 f^{*}(\bm{k}) & V_C^B + 2\gamma_2 \cos(k_z c) & 0 & 0\\
 0 & 0 & V_N & 0  \\
\gamma_1^B \left[1+\textrm{exp}(-\textrm{i}k_z c)\right] & 0 & 0 & V_B
\end{pmatrix},
\end{flalign}
\end{widetext}

where $V_C^A$, $V_C^B$, $V_B$, and $V_N$ are onsite potentials. The parameters $\gamma_i$, $i = \{0,1,2,5\}$ correspond to intra- and interlayer hopping amplitudes \footnote{Note that in this part of the appendix only, we explicitly write all the indices, i.e., $\gamma_{1}^B\equiv\gamma_B$}, \mbox{$f(\bm{k})=\textrm{e}^{\textrm{i}\tfrac{a}{\sqrt{3}}k_y}\bigl[1 + 2\,\textrm{e}^{-\textrm{i}\tfrac{\sqrt{3} a}{2} k_y}\cos\bigl({\tfrac{a}{2}k_x}\bigr) \bigr]\,,$} is the structural function of graphene, and $c = 2d$ is the lattice constant in $z$ direction. 
In analogy, the model Hamiltonian for the geometry in Fig.~\ref{Structure_parallel}(b) in the ordered basis (C$_{\textrm{A}}$, C$_{\textrm{B}}$, B, N) is

\begin{widetext}
\begin{flalign}
\mathcal{H}_{\textrm{p}}^{\textrm{(b)}} = & \begin{pmatrix}
V_C^A + 2\gamma_5 \cos(k_z c) & \gamma_0 f(\bm{k}) & 0 & \gamma_1^N \left[1+\textrm{exp}(\textrm{i}k_z c)\right]  \\
\gamma_0 f^{*}(\bm{k}) & V_C^B + 2\gamma_2 \cos(k_z c) & 0 & 0\\
 0 & 0 & V_B & 0  \\
\gamma_1^N \left[1+\textrm{exp}(-\textrm{i}k_z c)\right] & 0 & 0 & V_N
\end{pmatrix}.
\end{flalign}
\end{widetext}

Finally, the model Hamiltonian for the geometry in Fig.~\ref{Structure_parallel}(c) in the ordered basis (C$_{\textrm{A}}$, C$_{\textrm{B}}$, N, B) is

\begin{widetext}
\begin{flalign}
\mathcal{H}_{\textrm{p}}^{\textrm{(c)}} = & \begin{pmatrix}
V_C^A + 2\gamma_5 \cos(k_z c) & \gamma_0 f(\bm{k}) & 0 & \gamma_1^B \left[1+\textrm{exp}(\textrm{i}k_z c)\right]  \\
\gamma_0 f^{*}(\bm{k}) & V_C^B + 2\gamma_2 \cos(k_z c) & \gamma_1^N \left[1+\textrm{exp}(\textrm{i}k_z c)\right] & 0\\
 0 & \gamma_1^N \left[1+\textrm{exp}(-\textrm{i}k_z c)\right] & V_N & 0  \\
\gamma_1^B \left[1+\textrm{exp}(-\textrm{i}k_z c)\right] & 0 & 0 & V_B
\end{pmatrix},
\end{flalign}
\end{widetext}

\subsection{Results}
\begin{figure*}[htb]
     \includegraphics[width=0.8\linewidth]{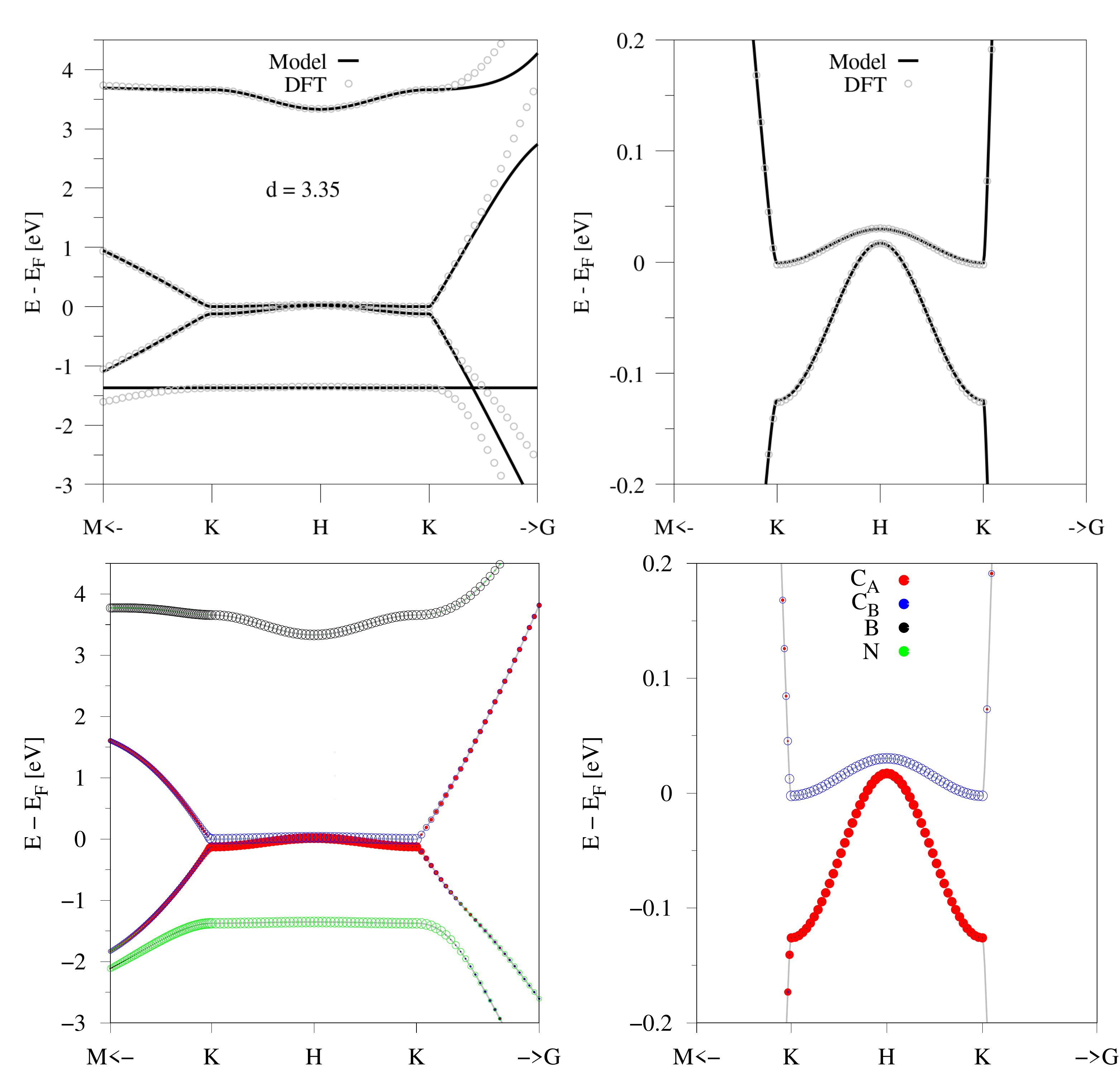}
     \caption{Top: DFT and fit results for the geometry in Fig.~\ref{Structure_parallel}(a). Bottom: The corresponding DFT-calculated projected band structure.
     }\label{fit_C_over_B}
    \end{figure*}

In Fig.~\ref{fit_C_over_B}, we provide the DFT and fit results for the 
geometry in Fig.~\ref{Structure_parallel}(a) for an interlayer distance of 3.35~\AA. The minimal model Hamiltonian Eq.~\eqref{DFT_Ham} accurately reproduces the low energy bands, with the fit parameters summarized in Table~\ref{tab:fit_distancea}. We also provide a projected band structure, to identify the sublattice contributions to the bands. 
In fact, the DFT results confirm that the energetically most favorable stacking results in a trivial (magenta) phase in the diagram of Fig.~\ref{fig:1}. 

\begin{figure*}[htb]
     \includegraphics[width=0.8\linewidth]{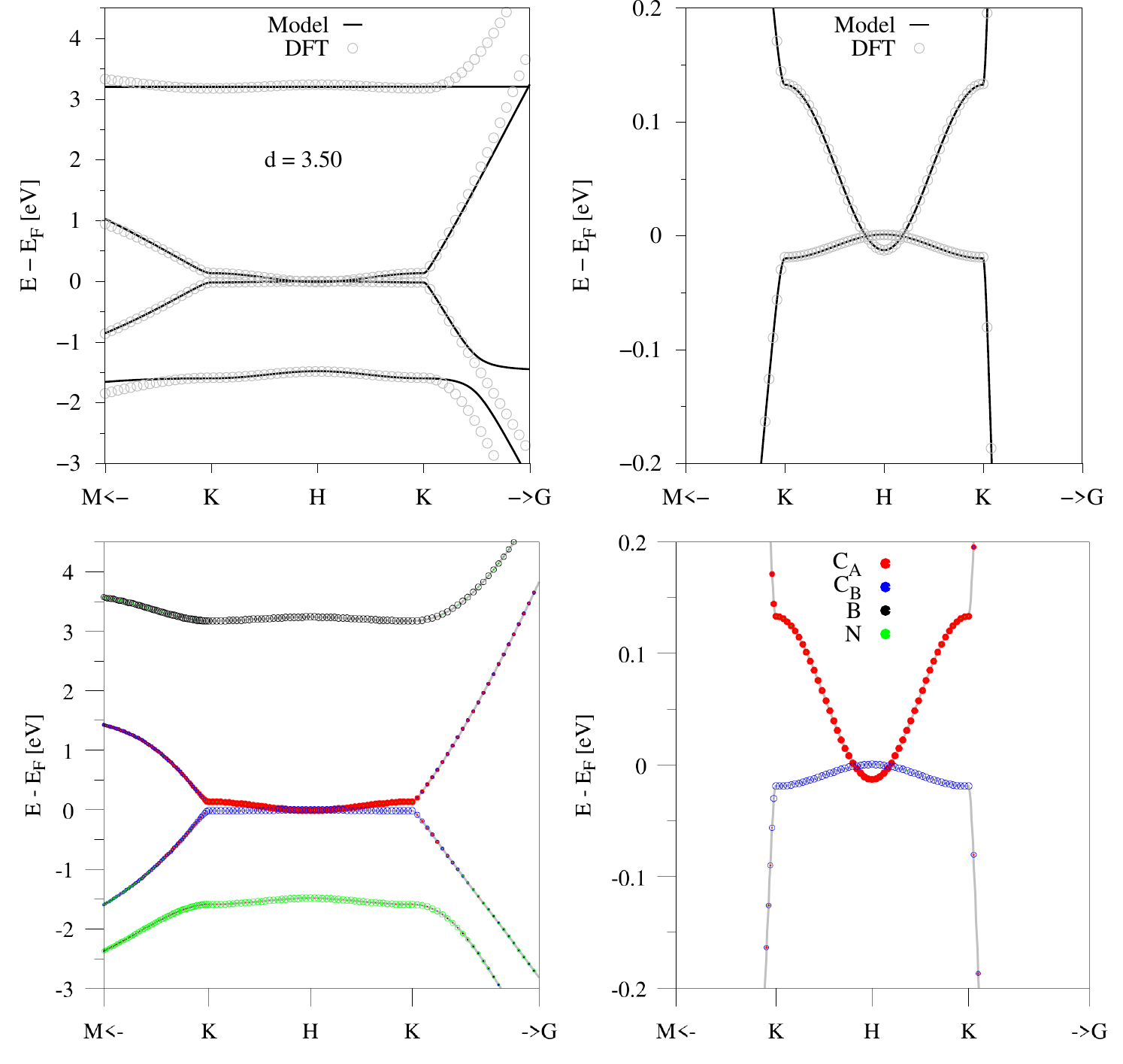}
     \caption{Top: DFT and fit results for the geometry in Fig.~\ref{Structure_parallel}(b). Bottom: The corresponding DFT-calculated projected band structure.
     }\label{fit_C_over_N}
    \end{figure*}

\begin{figure*}[htb]
     \includegraphics[width=0.8\linewidth]{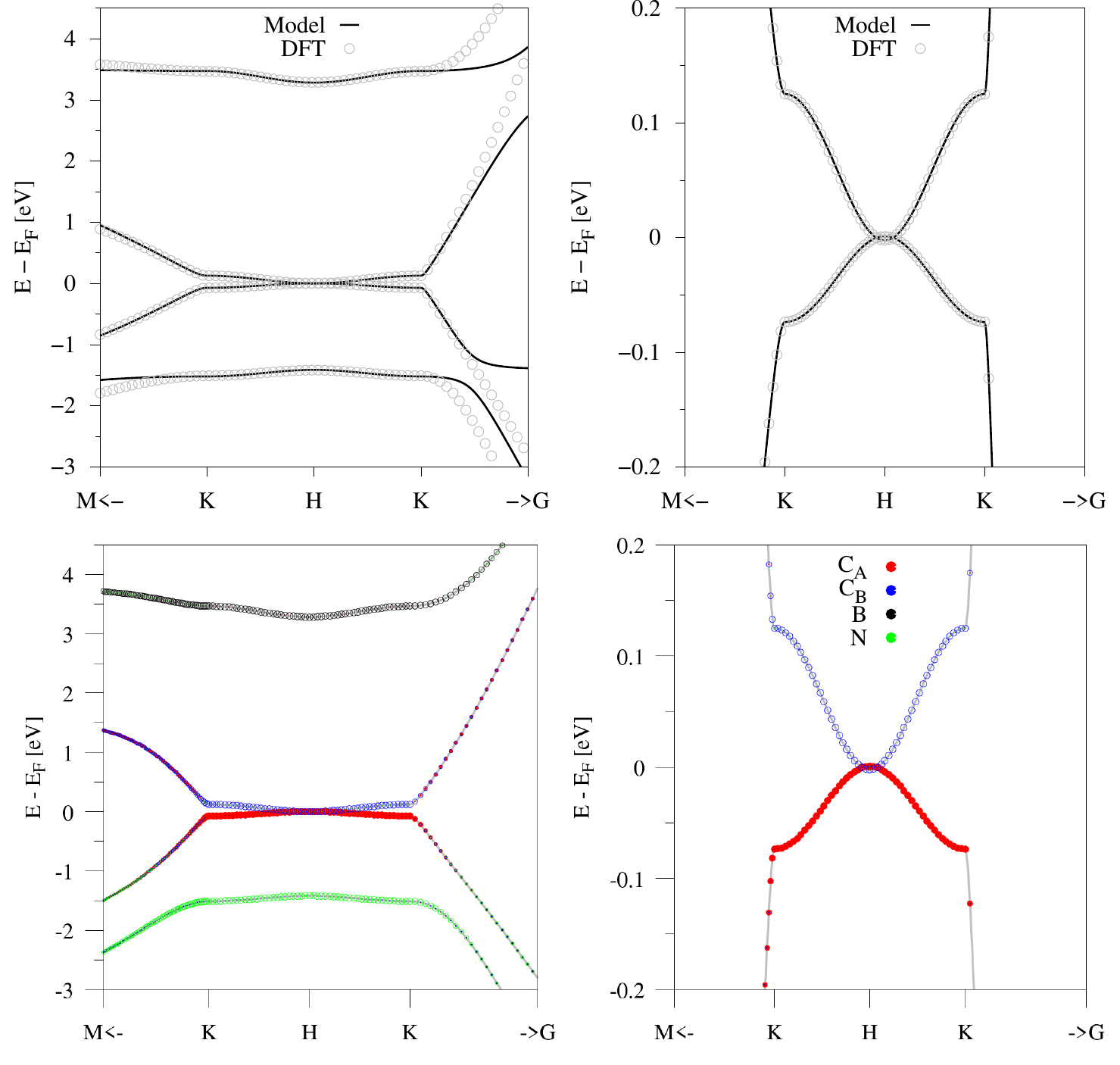}
     \caption{Top: DFT and fit results for the geometry in Fig.~\ref{Structure_parallel}(c). Bottom: The corresponding DFT-calculated projected band structure.
     }\label{fit_CC_over_BN}
    \end{figure*}

Similarly, we provide the DFT-calculated band structures with fit results for the other geometries in Fig.~\ref{fit_C_over_N} and Fig.~\ref{fit_CC_over_BN}, with the model parameters summarized in Table~\ref{tab:fit_distanceb} and Table~\ref{tab:fit_distancec}.
We find that the other stacking configurations provide low energy bands in the non-trivial phase. 
So it is likely that within a Gr/hBN moir\'e structure, both phases coexist. 

Finally, the interlayer distance dependence of the model parameters, for all three stacking configurations, is summarized in Fig.~\ref{Fig:distance}. In the Supplemental Material we also provide movies of the band structure evolution with the interlayer distance, demonstrating that a phase transition is not possible by tuning the distance. 

\begin{table*}[htb]
\caption{\label{tab:fit_distancea} The fit parameters of the model Hamiltonian $\mathcal{H}_{\textrm{p}}^{\textrm{(a)}}$, for different interlayer distances. The reference energy is the lowest energy. }
\begin{ruledtabular}
\begin{tabular}{cccccccccc}
d [\AA] &  $\gamma_0$ [eV] & $\gamma_1^B$ [eV] & $\gamma_2$ [eV] & $\gamma_5$ [eV] & $V_C^A$ [eV] & $V_C^B$ [eV] & $V_B$ [eV] & $V_N$ [eV] & $E - E_{\textrm{ref}}$ [eV] \\
\hline 
3.05 & 2.7547 & 0.826141 & -0.0210211 & 0.0961844 & 0.243162 & 0.0331826 & 3.3788 & -1.29091 & 0.00193936 \\
3.15 & 2.69242 & 0.7147 & -0.0150067 & 0.0763636 & 0.186408 & 0.0237476 & 3.35765 & -1.32144 & 0 \\
3.25 & 2.64012 & 0.617706 & -0.010691 & 0.0601839 & 0.145865 & 0.0203515 & 3.34286 & -1.34743 & 0.00832764 \\
3.35 & 2.59775 & 0.533638 & -0.00760797 & 0.0471555 & 0.11139 & 0.0145595 & 3.32984 & -1.37067 & 0.0221912 \\
3.45 & 2.564 & 0.460935 & -0.00541895 & 0.0367868 & 0.084494 & 0.00977721 & 3.31976 & -1.39077 & 0.0387604 \\
3.55 & 2.53771 & 0.398114 & -0.00388218 & 0.0285891 & 0.0639864 & 0.00630862 & 3.31176 & -1.40743 & 0.0562206 \\
3.65 & 2.51729 & 0.344088 & -0.00280525 & 0.0221904 & 0.0491791 & 0.0044938 & 3.305 & -1.42199 & 0.0735439 \\
3.75 & 2.50181 & 0.297844 & -0.00204633 & 0.0172377 & 0.0392811 & 0.00441093 & 3.3007 & -1.43176 & 0.0900766 \\
\end{tabular}
\end{ruledtabular}
\end{table*}

\begin{table*}[htb]
\caption{\label{tab:fit_distanceb} The fit parameters of the model Hamiltonian $\mathcal{H}_{\textrm{p}}^{\textrm{(b)}}$, for different interlayer distances.  }
\begin{ruledtabular}
\begin{tabular}{cccccccccc}
d [\AA] &  $\gamma_0$ [eV] & $\gamma_1^N$ [eV] & $\gamma_2$ [eV] & $\gamma_5$ [eV] & $V_C^A$ [eV] & $V_C^B$ [eV] & $V_B$ [eV] & $V_N$ [eV] & $E - E_{\textrm{ref}}$ [eV] \\
\hline 
3.10 & 2.25211 & 0.435216 & -0.0197129 & 0.0283977 & 0.0137979 & -0.0339898 & 3.0174 & -1.55462 & 0.069945 \\
3.20 & 2.29749 & 0.365096 & -0.0141096 & 0.0210104 & 0.0107601 & -0.0242352 & 3.08254 & -1.52662 & 0.0567718 \\
3.30 & 2.33659 & 0.306697 & -0.0101238 & 0.0150916 & 0.00725264 & -0.0175111 & 3.13329 & -1.50618 & 0.0548503 \\
3.40 & 2.36844 & 0.258148 & -0.00728036 & 0.0104537 & 0.00387735 & -0.0128508 & 3.17258 & -1.49182 & 0.0598679 \\
3.50 & 2.39339 & 0.217895 & -0.00524873 & 0.0069222 & 0.00112994 & -0.00940942 & 3.20297 & -1.48154 & 0.0691858 \\
3.60 & 2.41262 & 0.184943 & -0.0038015 & 0.00420361 & -0.00162581 & -0.00738836 & 3.22568 & -1.4747 & 0.0806541 \\
3.70 & 2.42711 & 0.158164 & -0.00279051 & 0.00212687 & -0.00354894 & -0.00567825 & 3.24278 & -1.47029 & 0.0932816 \\
3.80 & 2.43767 & 0.136573 & -0.00206618 & 0.00060251 & -0.00482932 & -0.00435118 & 3.25588 & -1.46734 & 0.106155 \\
\end{tabular}
\end{ruledtabular}
\end{table*}

\begin{table*}[htb]
\caption{\label{tab:fit_distancec} The fit parameters of the model Hamiltonian $\mathcal{H}_{\textrm{p}}^{\textrm{(c)}}$, for different interlayer distances. }
\begin{ruledtabular}
\begin{tabular}{ccccccccccc}
d [\AA] &  $\gamma_0$ [eV] & $\gamma_1^B$ [eV] & $\gamma_1^N$ [eV] & $\gamma_2$ [eV] & $\gamma_5$ [eV] & $V_C^A$ [eV] & $V_C^B$ [eV] & $V_B$ [eV] & $V_N$ [eV] & $E - E_{\textrm{ref}}$ [eV] \\
\hline 
3.15 & 1.98206 & 0.716265 & 0.403857 & 0.0235745 & 0.0778035 & 0.163119 & 0.0323085 & 3.25605 & -1.36919 & 0.0856805 \\
3.25 & 2.13136 & 0.619404 & 0.338775 & 0.0172232 & 0.0611696 & 0.126637 & 0.0246108 & 3.26689 & -1.38102 & 0.0705259 \\
3.35 & 2.23663 & 0.53526 & 0.284577 & 0.0121866 & 0.0478066 & 0.0983562 & 0.0182863 & 3.27438 & -1.39315 & 0.0670959 \\
3.45 & 2.30986 & 0.462367 & 0.239647 & 0.00828693 & 0.0372113 & 0.075884 & 0.0125553 & 3.27816 & -1.40536 & 0.0709821 \\
3.55 & 2.36046 & 0.399392 & 0.202657 & 0.00526414 & 0.0288518 & 0.0587337 & 0.00794564 & 3.28132 & -1.41596 & 0.0792004 \\
3.65 & 2.39517 & 0.345237 & 0.172457 & 0.00294773 & 0.0223473 & 0.0458979 & 0.00457288 & 3.2835 & -1.42555 & 0.0898538 \\
3.75 & 2.41887 & 0.298844 & 0.147998 & 0.00123478 & 0.0173249 & 0.0362649 & 0.00218736 & 3.28483 & -1.43317 & 0.101616 \\
3.85 & 2.43498 & 0.259298 & 0.128434 & -5.0675e-05 & 0.0134694 & 0.0286455 & 5.0704e-05 & 3.28621 & -1.43986 & 0.11372 \\
\end{tabular}
\end{ruledtabular}
\end{table*}

    \begin{figure*}[htb]
     \includegraphics[width=1\linewidth]{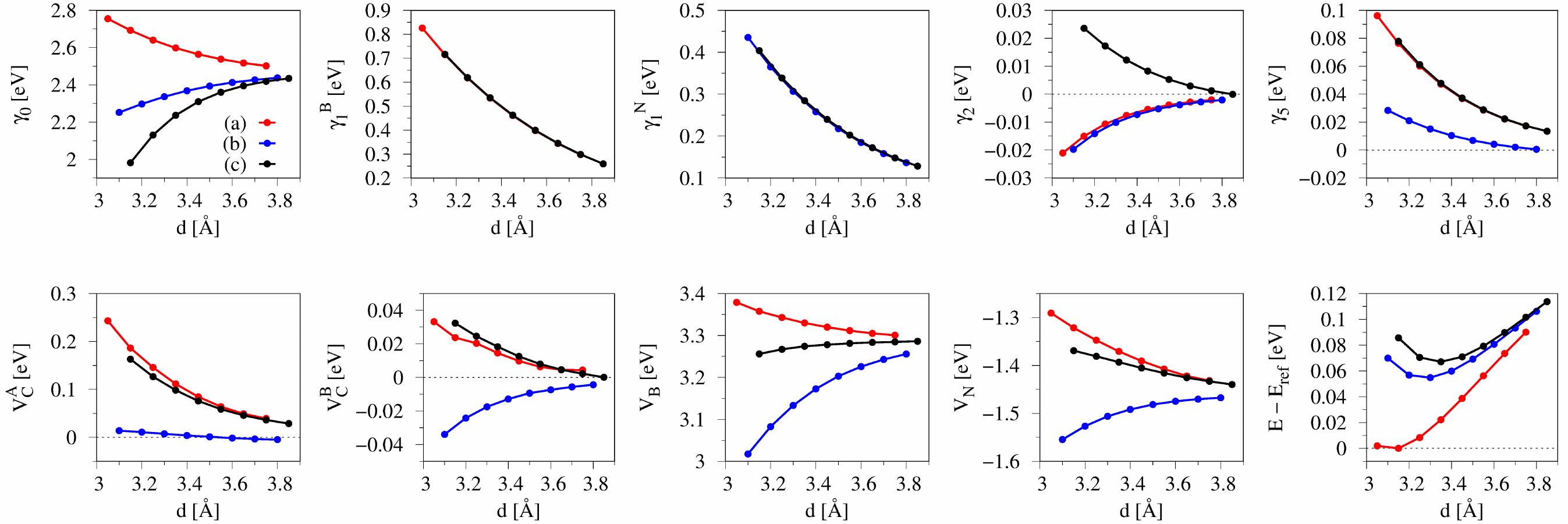}
     \caption{Evolution of the fit parameters as function of the interlayer distance, summarizing the results from Table~\ref{tab:fit_distancea}, Table~\ref{tab:fit_distanceb}, and Table~\ref{tab:fit_distancec}.
     }\label{Fig:distance}
    \end{figure*}

\clearpage
\bibliography{GrHBN3D}

\end{document}